\documentclass[aip,prb,showpacs,twocolumn]{revtex4}

\usepackage{graphics,graphicx,amssymb,amsmath,color}
\begin{document}

\title{Barlowite: A Spin-1/2 Antiferromagnet with a Geometrically Perfect Kagome Motif}
\author{Tian-Heng~Han$^{1,2\dag}$}
\author{John ~Singleton$^{3}$}
\author{John A.~Schlueter$^{2,4}$}
\affiliation{$^{1}$James Frank Institute and Department of Physics, University of Chicago, Chicago, Illinois 60637, USA}
\affiliation{$^{2}$Materials Science Division, Argonne National Laboratory, Argonne, Illinois 60439, USA}
\affiliation{$^{3}$National High Magnetic Field Laboratory, Los Alamos National Laboratory, Los Alamos, New Mexico 87545, USA}
\affiliation{$^{4}$Division of Materials Research, National Science Foundation, Arlington, Virginia 22230, USA}
\date{\today}

\begin{abstract}
We present thermodynamic studies of a new spin-1/2 antiferromagnet containing undistorted kagome lattices---barlowite Cu$_{4}$(OH)$_{6}$FBr.  Magnetic susceptibility gives $\theta_{CW}$ = $-$136 K, while long-range order does not happen until $T_{N}$ = 15 K with a weak ferromagnetic moment $\mu$ $<$ 0.1$\mu_{B}$/Cu.  A 60 T magnetic field induces a moment less than 0.5$\mu_{B}$/Cu at $T$ = 0.6 K.  Specific-heat measurements have observed multiple phase transitions at $T \ll$ $\mid$$\theta_{CW}$$\mid$.  The magnetic entropy of these transitions is merely 18\% of $k_{B}$ln2 per Cu spin.  These observations suggest that nontrivial spin textures are realized in barlowite with magnetic frustration.  Comparing with the leading spin-liquid candidate herbertsmithite, the superior interkagome environment of barlowite sheds light on new spin-liquid compounds with minimum disorder.  The robust perfect geometry of the kagome lattice makes charge doping promising.
\end{abstract}

\pacs{75.50Ee, 75.10Kt, 75.40Cx, 81.10.Dn} \maketitle

Magnetic materials are pervasive in modern physics and quantum mechanics produces unexpected behaviors.  A variety of exotic states---many of which are topological---can be hosted in quantum magnets with competing interactions.  Knowledge of unconventional magnetism in new compounds has great appeal across boundaries in physics.  One of the controversies is whether a quantum spin liquid (QSL), such as a resonating valence bond state\cite{Anderson1973}, can be realized experimentally.  In a QSL, all of the spins result in a long-range quantum entanglement and remain in motion even at a temperature of absolute zero\cite{Balents}.  A QSL cannot be described by broken symmetries in the same way as conventional magnets, and it represents new states of matter.  Having a zoo of exotic phenomena and being a potential key ingredient of high-$T_{c}$ superconductivity, experimental realizations of a QSL state have been a long and challenging pursuit for decades\cite{Anderson1987,Lee2008}.  The difficulty is rooted in precisely balancing the microscopic interactions with quantum fluctuations, which together prevent the spins from long-range ordering.  Two-dimensional $S$ = 1/2 lattices with geometric frustration---where all exchanges cannot be satisfied---are one of the promising protocols.  Almost all candidates order magnetically---because of nonstoichiometry issues, imperfect lattice geometries, large spins, or perturbing interactions---at low temperatures\cite{Ramirez,YoshidaM,Wills,Zorko2013}, even though their ordered spin textures are complicated and fascinating by themselves.

The leading candidate is the $x$ = 1 end member of Zn-paratacamite [Zn$_{x}$Cu$_{4-x}$(OH)$_{6}$Cl$_{2}$ with $x$ $>$ 1/3] called herbertsmithite\cite{Han3}.  This compound features a geometrically perfect $S$ = 1/2 Cu-kagome lattice with a dominating nearest-neighbor Heisenberg antiferromagnetic exchange.  As $x$ approaches 1, Cu$^{2+}$ ions on interkagome sites are replaced by nonmagnetic Zn$^{2+}$ ions.  At $x$ = 1, the Cu-kagome layers become a two-dimensional magnetic system.  Theoretically, the ground state of such a Heisenberg model can be a gapped QSL\cite{white} as well as a gapless one\cite{Iqbal1,Iqbal2}.  Experimentally, a spinon continuum$-$fractionalized spin excitations resulting from spin-charge separation$-$has been observed by neutron scattering, indicating a QSL ground state\cite{Han3}.  However, a precise determination of the stoichiometry of the nominal $x$ = 1 sample gives $x$ = 0.85\cite{Freedman}.  The excess Cu$^{2+}$ ions on the interkagome sites weakly couple to the kagome spins through Cu-O-Cu superexchange interactions.  This provides additional terms in the spin Hamiltonian and presents challenges to theoretical modeling.  In particular, the precise spin Hamiltonian of herbertsmithite remains ambiguous because of the infeasibility for spin wave study.  Alternative investigations on clinoatacamite---x = 0 mother compound of herbertsmithite with a magnetically ordered ground state---are devalued due to its Jahn-Teller distorted kagome structure---a ubiquitous conundrum for lattices with spin-1/2 transition metals.  At low temperatures or energies, the excess spins have strong response which overwhelms additional evidence of a QSL---the existence or absence of a spin gap\cite{white,Iqbal1,Iqbal2}.  In addition to the interkagome impurity, a trace amount of Zn$^{2+}$ ions in the Cu-kagome layer remains a nagging concern since these two 3$d$ transition metals are next to each other in the periodic table, though the Zn dilution is measured to be no more than $\sim$1\%\cite{Freedman}.  Doping the interkagome sites by large nonmagnetic Cd$^{2+}$ ions---a 4$d$ transition metal and difficult to exchange sites with Cu---unfortunately distorts the kagome structure and induces spin ordering\cite{McQueen}.  For the search of stronger evidence of a QSL, a new family of $S$ = 1/2 antiferromagnets featuring undistorted kagome lattces is all the more urgent.

\begin{figure}
\centering
\includegraphics[width=7.5cm]{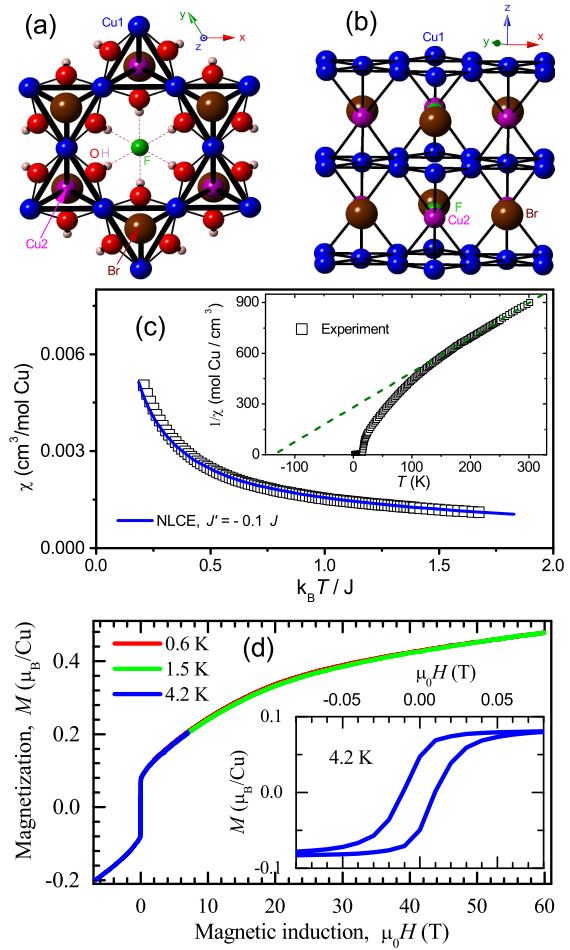} \vspace{-3mm}
\caption{(color online) Crystal structure of barlowite when looking perpendicular (a) and almost parallel (b) to the kagome lattice. The symbols are kagome Cu1-blue; interkagome Cu2-purple; O-red; H-salmon; F-green; Br-brown. In (a), the thick black lines denote Cu-Cu grids.  (c) Magnetic susceptibility of polycrystalline (explained in the text) barlowite measured at $\mu_{0}H$ = 0.1 T.  The data are compared with calculations\cite{Rigol} as described in the text. Inset: Inverse susceptibility and a Curie-Weiss fit.  (d) Magnetization versus field measured at liquid-helium temperatures.  Inset: Hysteresis loop at $\mu_{0}H$ $<$ 0.08 T.} \vspace{-6mm}
\label{InverseChi}
\end{figure}

Here we present a new candidate compound, barlowite Cu$_{4}$(OH)$_{6}$FBr\cite{Schlueter}, with its bulk properties studied using thermodynamic techniques.  Barlowite has a hexagonal crystal system in the $P$6$_{3}$/$mmc$ space group [$a$ = 6.6786(2) \AA, $c$ = 9.2744(3) \AA]\cite{BLW}.  As shown in Figs.~\ref{InverseChi}(a) and ~\ref{InverseChi}(b), three Cu$^{2+}$ ions in the formula are crystallographically equivalent and form a geometrically perfect kagome lattice.  The space of the interkagome site is so large that the fourth Cu$^{2+}$ ion sits in one of three equivalent positions (only the average position is shown).  The kagome layers stack on top of each other, different from the staggered stacking in Zn-paratacamite.  Barlowite orders magnetically at $T_{N}$ = 15 K and frustrated antiferromagnetism is present with multiple phase transitions at low temperatures. Doping the interlayer sites with large nonmagnetic ions is likely to succeed, as has been demonstrated in Zn-paratacamite.  The kagome spin lattices in barlowite are weakly coupled.  As a new mother compound of QSL states, the uniqueness and advantages of barlowite are discussed.

The sample was grown hydrothermally and was characterized by x-ray diffraction\cite{Schlueter}.  Magnetic susceptibility ($\chi\approx M/H$ in the paramagnetic regime and in the weak-field limit) as a function of temperature has been measured by using a SQUID magnetometer (Quantum Design MPMS) on a 68.5 mg polycrystalline sample---a collection of numerous small crystals.  In the inset of Fig.~\ref{InverseChi}(c), the inverse susceptibility is fitted with a Curie-Weiss function for 180 $< T <$ 300 K.  A temperature independent contribution, possibly from the core diamagnetism and the Van Vleck paramagnetism of the sample and the holder, has been subtracted.  The Curie-Weiss temperature is $\theta_{CW}$ = $-$136 $\pm$ 10 K, indicating strong antiferromagnetic exchange.  The mean-field $g$ factor of the Cu$^{2+}$ ions is 2.27, assuming $S$ = 1/2.

\begin{table}
\vspace{-2mm}
\caption{Bond angles of the superexchanges in Cu$_{4}$(OH)$_{6}$FBr at room temperature.  Cu1 is in the kagome plane, and Cu2 denotes the average position of three equivalent interkagome sites: Cu2a, Cu2b, and Cu2c.}
\vspace{3mm}
\begin{tabular}{| c | c | c | c | c |}
\hline
 &Cu1-O-Cu1 & Cu1-O-Cu2 & Cu1-O-Cu2a,b & Cu1-O-Cu2c\\
\hline
Angle & 117.4$^{\circ}$ & 95.8$^{\circ}$ & 88.7$^{\circ}$ & 107.5$^{\circ}$ \\
\hline
\end{tabular}
\vspace{-3mm}
\label{CuOCu}
\end{table}

Magnetization as a function of field was measured by using an extraction magnetometer\cite{Goddard'} in a $^{3}$He cryostat.  Magnetic fields up to 60 T were provided by a 25-millisecond-duration pulsed magnet at National High Magnetic Field Laboratory at Los Alamos\cite{Goddard'}.  The magnetometer was calibrated against SQUID measurements, as shown in Fig.~\ref{InverseChi}(d). The low-field behavior is dominated by a hysteresis loop with a coercive field of about 0.01 T (inset).  This weak ferromagnetism may be due to the interkagome Cu or Dzyaloshinskii-Moriya interaction (DMI).  The magnetization jump is less than 0.1$\mu_{B}$/Cu, showing that only a small fraction of the available magnetic moment is involved.  Above the hysteresis loop, $M$($H$) is monotonic.  The polarized moment is 0.48(4)$\mu_{B}$/Cu at 60 T, corroborating the strong antiferromagnetic exchange.  No impurity contribution is observed.

Magnetically, barlowite is better modeled by a stack of weakly coupled kagome layers---instead of a three-dimensional network of tetrahedrons---regarding $J'/J$, where $J'$ is the exchange between a kagome Cu and the average position for an interkagome Cu.  Down to $T$ $\sim$ 0.2$J/k_{B}$, the magnetic susceptibility has been calculated, using numerical linked-cluster expansion (NLCE), on a 16-site cluster by considering a kagome lattice coupled to interkagome spins\cite{Rigol}.  As shown in Fig.~\ref{InverseChi}(c), our susceptibility data are well described by assuming $J'/J$ = $-$0.1 with $J/k_{B}$ = $-$180 K---slightly less than $\theta_{CW}$ = $-$136 K.  In a frustrated magnet, a Curie-Weiss fit in a temperature range comparable to $\mid$$\theta_{CW}$$\mid$ often needs to be corrected when quantifying the microscopic exchanges.  In addition, $\theta_{CW}$ describes the combined effect of all exchanges.  As shown in Table.~\ref{CuOCu}, $J$ and $J'$ of barlowite are consistent with the Goodenough-Kanamori-Anderson rule for the Cu-O-Cu superexchange bond angles\cite{Mizuno}. As has been demonstrated in a metal-organic kagome compound, competing antiferromagnetic and ferromagnetic interactions result in a quick saturation of the Cu spin moments at $\mu_{B} B \sim$ $J$/20\cite{Nytko}.  In herbertsmithite, where most of the interkagome sites are occupied by nonmagnetic Zn$^{2+}$ ions, antiferromagnetic exchange in the kagome lattice dominates, and only 0.1$\mu_{B}$/Cu is induced at $\mu_{B} B \sim$ $J$/3\cite{Han4}. Barlowite falls in between with 0.48$\mu_{B}$/Cu induced at $\mu_{B} B \sim$ $J$/2---consistent with a combined effect from strong in-kagome antiferromagnetic and weak out-of-kagome ferromagnetic exchanges.

\begin{figure}
\centering
\includegraphics[width=7.7cm]{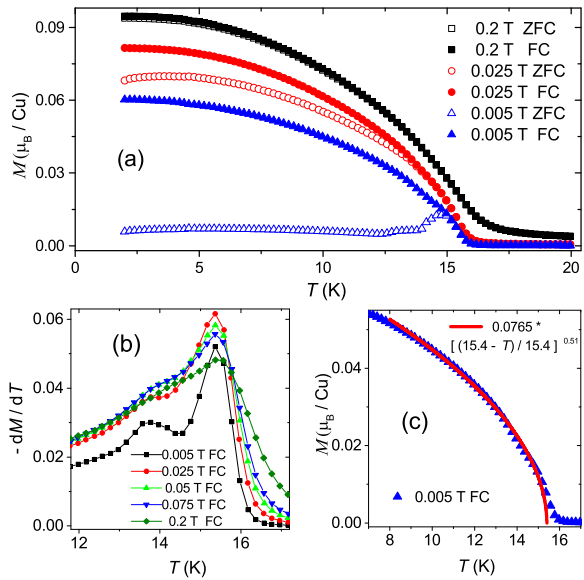} \vspace{-3mm}
\caption{(color online) (a) Magnetization versus temperature measured after cooling from room temperature in zero field (ZFC) and a field (FC).  (b) Negative derivative of the temperature dependence of magnetization in the proximity of $T_{N}$.  (c) Power-law fitting to the spontaneous magnetization.} \vspace{-6mm}
\label{ZFCFC}
\end{figure}

Magnetization at low fields is plotted in Fig.~\ref{ZFCFC}(a), showing a phase transition to a long-range ordered state with a small ferromagnetic moment at N\'{e}el temperature $T_{N}$ = 15.4 K.  At $\mu_{0}H$ = 0.005 T, zero-field-cooled (ZFC) and field-cooled (FC) data indicate a thermal hysteresis, which gradually vanishes as field increases beyond the coercive field.  The thermal hysteresis may originate from domains of the weak ferromagnetism or a trace amount of spin-glass phase, neither of which plays a major role since ZFC and FC curves collapse onto each other at $\mu_{0}H$ = 0.2 T.  In Fig.~\ref{ZFCFC}(b), we have plotted the negative derivatives of the FC data shown in Fig.~\ref{ZFCFC}(a) in order to precisely detect phase transitions.  At $\mu_{0}H$ = 0.005 T, a second phase transition occurs at $T$ = 13.8 K, which broadens at increasing fields and becomes a shoulder of the main peak at 0.2 T.  In Fig.~\ref{ZFCFC}(c), the spontaneous magnetization below $T_{N}$ is fitted to a power law $M$ = $At^{\beta}$ giving $\beta$ = 0.51, where $t$ = $\mid$$T$ $-$ $T_{N}\mid$/$T_{N}$ is the reduced temperature and $A$ is a constant.  Similar fits at fields up to 0.2 T give $\beta$ between 0.44 and 0.51.  This exponent is larger than 0.39 observed for the three-dimensional ferromagnetic ordering of iron\cite{Febeta}.  At $\mu_{0}H$ = 0.005 T, the ordered moment saturates at 0.06$\mu_{B}$/Cu in the $T$ $\rightarrow$ 0 limit, as shown by the FC curve in Fig.~\ref{ZFCFC}(a).

\begin{figure}
\centering
\includegraphics[width=7.5cm]{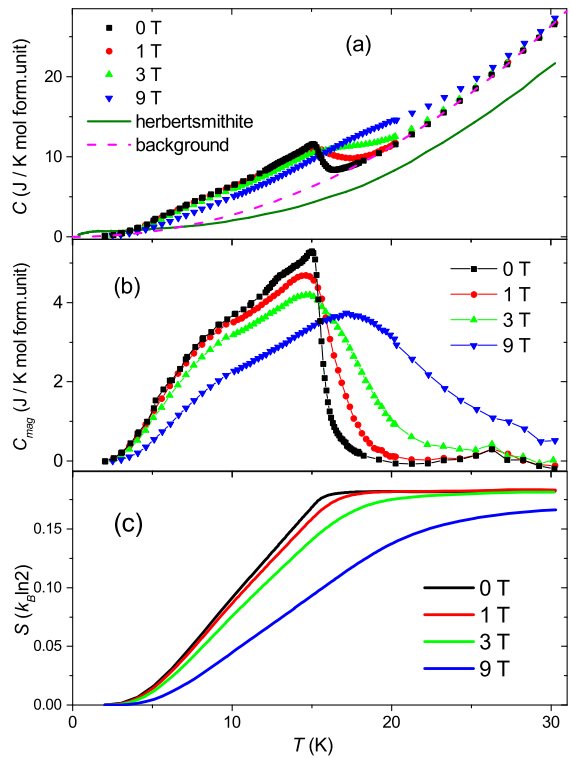} \vspace{-3mm}
\caption{(color online) (a) Specific heat measured on a single crystal sample of barlowite at fields parallel to the kagome lattice and compared with zero-field specific heat of herbertsmithite.  (b) Magnetic specific heat after subtracting the background.  (c)  Magnetic entropy integrated from $T$ = 2 K and normalized as a fraction of the total value per Cu spin.} \vspace{-6mm}
\label{Specificheat}
\end{figure}

Specific heat was measured using a Quantum Design physical property measurement system (PPMS) on a 5.4 mg single-crystal sample.  The field was applied parallel to the kagome plane.  In Fig.~\ref{Specificheat}(a), at zero field, a phase transition at $T$ = 15 K corroborates the magnetization measurements.  The application of a field progressively pushes the entropy below $T_{N}$ to higher temperatures, indicating that a large part of the low-temperature specific heat is magnetic.  A full suppression of the magnetic entropy requires $\mu_{0}H$ $>$ 20 T.  The background specific heat of phonon and disordered spins is estimated by fitting the zero-field specific heat between 20 and 30 K to a polynomial $C_{bg}$ = $aT^{2}$+$bT^{3}$.  The specific heat at $T$ $>$ 30 K deviates from a simple polynomial.  The returned parameters are $a$ = 0.0263 J/K$^{3}$ mol form. unit and $b$ = 9.87 $\times$ 10$^{-5}$ J/K$^{4}$ mol form. unit.  Instead of the $T^{3}$ law expected for a lattice structure in 3D, the $T^{2}$ term dominates $C_{bg}$, possibly due to spin correlations formed above $T_{N}$.  $C_{bg}$ resembles the total specific heat of polycrystalline herbertsmithite with disordered spins\cite{Helton'}.  The magnetic specific heat $C_{mag}$ is obtained by subtracting $C_{bg}$ and is shown in Fig. 3(b).  As the temperature approaches $T_{N}$ from above at zero field, it is difficult to determine the critical exponent $\alpha$ in $C_{mag}$ $\propto$ $t^{-\alpha}$, since its value depends on $C_{bg}$.  When cooling below $T_{N}$, instead of falling towards zero in a power law, a broad hump extends down to $T$ $<$ 5 K.  A small kink in $C_{mag}$ is seen at $T$ = 13.8 K, signaling a second phase transition.  For 5 $<$ $T$ $<$ 10 K, the dome in $C_{mag}$ might indicate slow freezing of the spin moments.  At $\mu_{0}H$ = 1 T, the two closely spaced transitions at $T_{N}$ and 13.8 K become one rounded peak,  which is consistent with the magnetization data in Fig.~\ref{ZFCFC}(b).  For $T$ $>$ $T_{N}$, a third phase transition, which is very weak and insensitive to a field of 3 T, takes place at $T$ = 26 K.  This phase transition has not been observed in magnetization measurements.

Obtained by integrating $C_{mag}/T$ from $T$ = 2 to 30 K, Fig.~\ref{Specificheat}(c) shows the magnetic entropy released from the spin ordering transitions.  The plateaus at high fields are artifacts from the estimation of $C_{bg}$.  At zero field, the integration gives merely 18\% of what is expected for the Cu spins.  There is no indication, down to the lowest temperature measured, that residual entropy exists.  Apart from the small ordered moments of spin-1/2 ions, the missing entropy may be ascribed to the formation of dynamic spin correlations, such as a chiral spin state\cite{Grohol} or a valence bond solid\cite{MatanVBS}, far above $T_{N}$.  Such a behavior is ubiquitous for geometrically frustrated magnets and is also indicated by the deviation from the Curie-Weiss fit at $T$ $<$ 180 K in Fig.~\ref{InverseChi}(c).  For barlowite, the empirical parameter of frustration $f$ = $\mid$$\theta_{CW}$$\mid$/$T_{N}$ = 9 $\gg$ 1, indicating the presence of strong frustration\cite{Ramirezf}.  Structurally, it is unclear whether and how the interkagome Cu$^{2+}$ ions freeze into one of the three equivalent positions.  Such a process releases an entropy of $k_{B}$ln3 per form. unit, 40\% of the total entropy from Cu spins (4$k_{B}$ln2 per form. unit).

DMI can cause spin canting in a kagome lattice, generating ferromagnetic moments. It also affects the magnetic ground state, possibly tuning it in the proximity of a putative quantum critical point\cite{Cepas}. In herbertsmithite, the OH bonds dangle and may freeze randomly at about 50 K, complicating the DMI and spin dynamics\cite{Imai}. Similarly, in barlowite, $J$ and $J'$ are mediated by OH bonds, and both the in-plane and out-of-plane components of DMI are allowed by symmetry.  However, the OH bonds are stabilized because all H$^{+}$ ions are connected to F$^{-}$ ions---through strong hydrogen bonds---between the kagome layers.  Since the spin exchanges are sensitive to the hydrogen positions, the HF bond makes possible an accurate determination of the superexchange.  It remains unclear how the spinon continuum of herbertsmithite is related to its proximity to a quantum critical point.  While resembling many aspects of clinoatacamite and herbertsmithite, the different structure of barlowite provides an opportunity to fine-tune the exchange parameters.

Studies on QSL physics largely rely on the reconciliation between experiments and theories\cite{Balents}.  This requires the knowledge of the spin Hamiltonian, which cannot be determined precisely for herbertsmithite in the absence of spin waves\cite{Han2}.  Studies on clinoatacamite may not be helpful since its structure deviates from that of herbertsmithite as a result of Jahn-Teller distortion\cite{MendelsJPSJ}.  In clinoatacamite, $\theta_{CW}$ is $-$190 K and $J'$\,$\sim$ $-$0.1\,$J$\cite{Rigol}.  Three magnetic transitions have been observed---two closely spaced at 6 K and one at 18 K\cite{XGZheng}.  A small ferromagnetic moment $\sim$0.06$\mu_{B}$/Cu is observed at $T$ $<$ 6 K and $\mu$SR measurements support the magnetic nature of the 18 K transition.  These properties are similar to those of barlowite.  Unfortunately, no neutron scattering experiment has been performed on a single crystal sample and the microscopic spin properties of clinoatacamite remain under debate.

Barlowite sheds light on solutions. Its high structural symmetry relieves crystallographic twinning and allows the growth of large single crystals\cite{Han}. Even though the interkagome sites are fully occupied by Cu, in contrast to clinoatacamite, barlowite maintains the perfect kagome motif.  For neutron scattering experiments, the absorption and incoherent cross sections of deuterated barlowite are much reduced from those of herbertsmithite.  This provides an advantage to resolve subtle features at very low energies as well as spin wave dispersions for accurate derivation of the spin Hamiltonian.

The long-range ordering of barlowite at $T_{N}$ may be suppressed by replacing the interkagome Cu with nonmagnetic ions.  The large space around the interkagome site allows many options of 4$d$ transition ions, such as Sn$^{2+}$ or Cd$^{2+}$ ions, paving the way for stoichiometric $M$Cu$_{3}$(OH)$_{6}$FBr ($M$ = Sn, Cd, etc) without a concern for antisite disorder.  Based on the structure of herbertsmithite, density functional theory calculation demonstrates the possibility of substituting the interkagome Cu with ions of different valence states, modifying the electronic band structure of the kagome lattice\cite{Mazin}.  For barlowite, the perfect geometry of the kagome lattice---a core ingredient of QSLs---is extremely robust, making charge doping a real possibility.

In conclusion, a new $S$ = 1/2 antiferromagnet containing geometrically perfect kagome layers is realized by barlowite Cu$_{4}$(OH)$_{6}$FBr with weak out-of-plane ferromagnetic exchanges.  Multiple magnetic phase transitions occur at temperatures much lower than the Curie-Weiss temperature, indicating strong frustration and nontrivial spin orders.  Single crystals are available for future neutron scattering experiments.  By replacing the interkagome Cu with nonmagnetic ions, new candidates of QSLs will likely emerge with minimum disorder.  Regarding the mechanism of high-temperature superconductivity\cite{PALee2006}, barlowite gives new hope for doping a resonating valence bond state.

We thank Patrick Lee, Yasu Takano and Michael Norman for careful reading of the manuscript with useful comments.  T.-H. H. thanks the support of the Grainger Fellowship provided by the Department of Physics, University of Chicago.  T.-H. H. and J. A. S. were supported by the Argonne National Laboratory under contract with the U.S. Department of Energy (DOE) (DE-AC02-06CH11357).  A portion of this work was performed at the National High Magnetic Field Laboratory, which is supported by National Science Foundation Cooperative Agreement No. DMR-1157490, the State of Florida, the U.S. DOE, and through the U.S. DOE Basic Energy Science Field Work Proposal ``Science in 100 T."  J. S. thanks the University of Oxford for support.  J. A. S. acknowledges support from the Independent Research and Development program while serving at the National Science Foundation.

$^{\dag}$tianheng@alum.mit.edu\\

\bibliography{BLW}

\begin{thebibliography}{34}
\expandafter\ifx\csname natexlab\endcsname\relax\def\natexlab#1{#1}\fi
\expandafter\ifx\csname bibnamefont\endcsname\relax
  \def\bibnamefont#1{#1}\fi
\expandafter\ifx\csname bibfnamefont\endcsname\relax
  \def\bibfnamefont#1{#1}\fi
\expandafter\ifx\csname citenamefont\endcsname\relax
  \def\citenamefont#1{#1}\fi
\expandafter\ifx\csname url\endcsname\relax
  \def\url#1{\texttt{#1}}\fi
\expandafter\ifx\csname urlprefix\endcsname\relax\def\urlprefix{URL }\fi
\providecommand{\bibinfo}[2]{#2}
\providecommand{\eprint}[2][]{\url{#2}}

\bibitem[{\citenamefont{Anderson}(1973)}]{Anderson1973}
\bibinfo{author}{\bibfnamefont{P.~W.} \bibnamefont{Anderson}},
  \bibinfo{journal}{Mater. Res. Bull.} \textbf{\bibinfo{volume}{8}},
  \bibinfo{pages}{153} (\bibinfo{year}{1973}).

\bibitem[{\citenamefont{Balents}(2010)}]{Balents}
\bibinfo{author}{\bibfnamefont{L.}~\bibnamefont{Balents}},
  \bibinfo{journal}{Nature (London)} \textbf{\bibinfo{volume}{464}},
  \bibinfo{pages}{199} (\bibinfo{year}{2010}).

\bibitem[{\citenamefont{Anderson}(1987)}]{Anderson1987}
\bibinfo{author}{\bibfnamefont{P.~W.} \bibnamefont{Anderson}},
  \bibinfo{journal}{Science} \textbf{\bibinfo{volume}{235}},
  \bibinfo{pages}{1196} (\bibinfo{year}{1987}).

\bibitem[{\citenamefont{Lee}(2008)}]{Lee2008}
\bibinfo{author}{\bibfnamefont{P.~A.} \bibnamefont{Lee}},
  \bibinfo{journal}{Science} \textbf{\bibinfo{volume}{321}},
  \bibinfo{pages}{1306} (\bibinfo{year}{2008}).

\bibitem[{\citenamefont{Ramirez et~al.}(1990)\citenamefont{Ramirez, Espinosa,
  and Cooper}}]{Ramirez}
\bibinfo{author}{\bibfnamefont{A.~P.} \bibnamefont{Ramirez}},
  \bibinfo{author}{\bibfnamefont{G.~P.} \bibnamefont{Espinosa}},
  \bibnamefont{and} \bibinfo{author}{\bibfnamefont{A.~S.}
  \bibnamefont{Cooper}}, \bibinfo{journal}{Phys. Rev. Lett.}
  \textbf{\bibinfo{volume}{64}}, \bibinfo{pages}{2070} (\bibinfo{year}{1990}).

\bibitem[{\citenamefont{Yoshida et~al.}(2009)\citenamefont{Yoshida, Takigawa,
  Yoshida, Okamoto, and Hiroi}}]{YoshidaM}
\bibinfo{author}{\bibfnamefont{M.}~\bibnamefont{Yoshida}},
  \bibinfo{author}{\bibfnamefont{M.}~\bibnamefont{Takigawa}},
  \bibinfo{author}{\bibfnamefont{H.}~\bibnamefont{Yoshida}},
  \bibinfo{author}{\bibfnamefont{Y.}~\bibnamefont{Okamoto}}, \bibnamefont{and}
  \bibinfo{author}{\bibfnamefont{Z.}~\bibnamefont{Hiroi}},
  \bibinfo{journal}{Phys. Rev. Lett.} \textbf{\bibinfo{volume}{103}},
  \bibinfo{pages}{077207} (\bibinfo{year}{2009}).

\bibitem[{\citenamefont{Wills}(2001)}]{Wills}
\bibinfo{author}{\bibfnamefont{A.~S.} \bibnamefont{Wills}},
  \bibinfo{journal}{Phys. Rev. B} \textbf{\bibinfo{volume}{63}},
  \bibinfo{pages}{064430} (\bibinfo{year}{2001}).

\bibitem[{\citenamefont{Zorko et~al.}(2013)\citenamefont{Zorko, Bert,
  Ozarowski, van Tol, Boldrin, Wills, and Mendels}}]{Zorko2013}
\bibinfo{author}{\bibfnamefont{A.}~\bibnamefont{Zorko}},
  \bibinfo{author}{\bibfnamefont{F.}~\bibnamefont{Bert}},
  \bibinfo{author}{\bibfnamefont{A.}~\bibnamefont{Ozarowski}},
  \bibinfo{author}{\bibfnamefont{J.}~\bibnamefont{van Tol}},
  \bibinfo{author}{\bibfnamefont{D.}~\bibnamefont{Boldrin}},
  \bibinfo{author}{\bibfnamefont{A.~S.} \bibnamefont{Wills}}, \bibnamefont{and}
  \bibinfo{author}{\bibfnamefont{P.}~\bibnamefont{Mendels}},
  \bibinfo{journal}{Phys. Rev. B} \textbf{\bibinfo{volume}{88}},
  \bibinfo{pages}{144419} (\bibinfo{year}{2013}).

\bibitem[{\citenamefont{Han et~al.}(2012{\natexlab{a}})\citenamefont{Han,
  Helton, Chu, Nocera, Rodrigues-Rivera, Broholm, and Lee}}]{Han3}
\bibinfo{author}{\bibfnamefont{T.-H.} \bibnamefont{Han}},
  \bibinfo{author}{\bibfnamefont{J.~S.} \bibnamefont{Helton}},
  \bibinfo{author}{\bibfnamefont{S.}~\bibnamefont{Chu}},
  \bibinfo{author}{\bibfnamefont{D.~G.} \bibnamefont{Nocera}},
  \bibinfo{author}{\bibfnamefont{J.~A.} \bibnamefont{Rodrigues-Rivera}},
  \bibinfo{author}{\bibfnamefont{C.}~\bibnamefont{Broholm}}, \bibnamefont{and}
  \bibinfo{author}{\bibfnamefont{Y.~S.} \bibnamefont{Lee}},
  \bibinfo{journal}{Nature (London)} \textbf{\bibinfo{volume}{492}},
  \bibinfo{pages}{406} (\bibinfo{year}{2012}{\natexlab{a}}).

\bibitem[{\citenamefont{Yan et~al.}(2011)\citenamefont{Yan, Huse, and
  White}}]{white}
\bibinfo{author}{\bibfnamefont{S.}~\bibnamefont{Yan}},
  \bibinfo{author}{\bibfnamefont{D.}~\bibnamefont{Huse}}, \bibnamefont{and}
  \bibinfo{author}{\bibfnamefont{S.}~\bibnamefont{White}},
  \bibinfo{journal}{Science} \textbf{\bibinfo{volume}{332}},
  \bibinfo{pages}{1173} (\bibinfo{year}{2011}).

\bibitem[{\citenamefont{Iqbal et~al.}(2011)\citenamefont{Iqbal, Becca, and
  Poilblanc}}]{Iqbal1}
\bibinfo{author}{\bibfnamefont{Y.}~\bibnamefont{Iqbal}},
  \bibinfo{author}{\bibfnamefont{F.}~\bibnamefont{Becca}}, \bibnamefont{and}
  \bibinfo{author}{\bibfnamefont{D.}~\bibnamefont{Poilblanc}},
  \bibinfo{journal}{Phys. Rev. B} \textbf{\bibinfo{volume}{84}},
  \bibinfo{pages}{020407R} (\bibinfo{year}{2011}).

\bibitem[{\citenamefont{Iqbal et~al.}(2014)\citenamefont{Iqbal, Poilblanc, and
  Becca}}]{Iqbal2}
\bibinfo{author}{\bibfnamefont{Y.}~\bibnamefont{Iqbal}},
  \bibinfo{author}{\bibfnamefont{D.}~\bibnamefont{Poilblanc}},
  \bibnamefont{and} \bibinfo{author}{\bibfnamefont{F.}~\bibnamefont{Becca}},
  \bibinfo{journal}{Phys. Rev. B} \textbf{\bibinfo{volume}{89}},
  \bibinfo{pages}{020407R} (\bibinfo{year}{2014}).

\bibitem[{\citenamefont{Freedman et~al.}(2010)\citenamefont{Freedman, Han,
  Prodi, M\"{u}ller, Huang, Chen, Webb, Lee, McQueen, and Nocera}}]{Freedman}
\bibinfo{author}{\bibfnamefont{D.~E.} \bibnamefont{Freedman}},
  \bibinfo{author}{\bibfnamefont{T.~H.} \bibnamefont{Han}},
  \bibinfo{author}{\bibfnamefont{A.}~\bibnamefont{Prodi}},
  \bibinfo{author}{\bibfnamefont{P.}~\bibnamefont{M\"{u}ller}},
  \bibinfo{author}{\bibfnamefont{Q.~Z.} \bibnamefont{Huang}},
  \bibinfo{author}{\bibfnamefont{Y.~S.} \bibnamefont{Chen}},
  \bibinfo{author}{\bibfnamefont{S.~M.} \bibnamefont{Webb}},
  \bibinfo{author}{\bibfnamefont{Y.~S.} \bibnamefont{Lee}},
  \bibinfo{author}{\bibfnamefont{T.~M.} \bibnamefont{McQueen}},
  \bibnamefont{and} \bibinfo{author}{\bibfnamefont{D.~G.}
  \bibnamefont{Nocera}}, \bibinfo{journal}{J. Am. Chem. Soc.}
  \textbf{\bibinfo{volume}{132}}, \bibinfo{pages}{16185}
  (\bibinfo{year}{2010}).

\bibitem[{\citenamefont{McQueen et~al.}(2011)\citenamefont{McQueen, Han,
  Freedman, Stephens, Lee, and Nocera}}]{McQueen}
\bibinfo{author}{\bibfnamefont{T.~M.} \bibnamefont{McQueen}},
  \bibinfo{author}{\bibfnamefont{T.~H.} \bibnamefont{Han}},
  \bibinfo{author}{\bibfnamefont{D.~E.} \bibnamefont{Freedman}},
  \bibinfo{author}{\bibfnamefont{P.~W.} \bibnamefont{Stephens}},
  \bibinfo{author}{\bibfnamefont{Y.~S.} \bibnamefont{Lee}}, \bibnamefont{and}
  \bibinfo{author}{\bibfnamefont{D.~G.} \bibnamefont{Nocera}},
  \bibinfo{journal}{J. Solid State Chem.} \textbf{\bibinfo{volume}{184}},
  \bibinfo{pages}{3319} (\bibinfo{year}{2011}).

\bibitem[{Sch()}]{Schlueter}
\bibinfo{note}{See Supplemental Material for more information.}

\bibitem[{\citenamefont{Elliott and Cooper}(2010)}]{BLW}
\bibinfo{author}{\bibfnamefont{P.}~\bibnamefont{Elliott}} \bibnamefont{and}
  \bibinfo{author}{\bibfnamefont{M.~A.} \bibnamefont{Cooper}},
  \bibinfo{journal}{Mineral. Mag.} \textbf{\bibinfo{volume}{74}},
  \bibinfo{pages}{798} (\bibinfo{year}{2010}).

\bibitem[{\citenamefont{Khatami et~al.}(2012)\citenamefont{Khatami, Helton, and
  Rigol}}]{Rigol}
\bibinfo{author}{\bibfnamefont{E.}~\bibnamefont{Khatami}},
  \bibinfo{author}{\bibfnamefont{J.~S.} \bibnamefont{Helton}},
  \bibnamefont{and} \bibinfo{author}{\bibfnamefont{M.}~\bibnamefont{Rigol}},
  \bibinfo{journal}{Phys. Rev. B} \textbf{\bibinfo{volume}{85}},
  \bibinfo{pages}{064401} (\bibinfo{year}{2012}).

\bibitem[{God()}]{Goddard'}
\bibinfo{note}{P. A. Goddard, T. Lancaster, S. J. Blundell, J. Singleton, P.
  Sengupta, R. D. McDonald, S. Cox, N. Harrison, F. L. Pratt, J. L. Manson, H.
  I. Southerland and J. A. Schlueter, New J. Phys. \textbf{10}, 083025,
  (2008)}.

\bibitem[{\citenamefont{Mizuno et~al.}(1998)\citenamefont{Mizuno, Tohyama,
  Maekawa, Osafune, Motoyama, Eisaki, and Uchida}}]{Mizuno}
\bibinfo{author}{\bibfnamefont{Y.}~\bibnamefont{Mizuno}},
  \bibinfo{author}{\bibfnamefont{T.}~\bibnamefont{Tohyama}},
  \bibinfo{author}{\bibfnamefont{S.}~\bibnamefont{Maekawa}},
  \bibinfo{author}{\bibfnamefont{T.}~\bibnamefont{Osafune}},
  \bibinfo{author}{\bibfnamefont{N.}~\bibnamefont{Motoyama}},
  \bibinfo{author}{\bibfnamefont{H.}~\bibnamefont{Eisaki}}, \bibnamefont{and}
  \bibinfo{author}{\bibfnamefont{S.}~\bibnamefont{Uchida}},
  \bibinfo{journal}{Phys. Rev. B} \textbf{\bibinfo{volume}{57}},
  \bibinfo{pages}{5326} (\bibinfo{year}{1998}).

\bibitem[{\citenamefont{Nytko et~al.}(2008)\citenamefont{Nytko, Helton,
  M\"{u}ller, and Nocera}}]{Nytko}
\bibinfo{author}{\bibfnamefont{E.~A.} \bibnamefont{Nytko}},
  \bibinfo{author}{\bibfnamefont{J.~S.} \bibnamefont{Helton}},
  \bibinfo{author}{\bibfnamefont{P.}~\bibnamefont{M\"{u}ller}},
  \bibnamefont{and} \bibinfo{author}{\bibfnamefont{D.~G.}
  \bibnamefont{Nocera}}, \bibinfo{journal}{J. Am. Chem. Soc.}
  \textbf{\bibinfo{volume}{130}}, \bibinfo{pages}{2922} (\bibinfo{year}{2008}).

\bibitem[{\citenamefont{Han et~al.}()\citenamefont{Han, Chisnell, Bonnoit,
  Freedman, Zapf, Harrison, Nocera, Takano, and Lee}}]{Han4}
\bibinfo{author}{\bibfnamefont{T.-H.} \bibnamefont{Han}},
  \bibinfo{author}{\bibfnamefont{R.}~\bibnamefont{Chisnell}},
  \bibinfo{author}{\bibfnamefont{C.~J.} \bibnamefont{Bonnoit}},
  \bibinfo{author}{\bibfnamefont{D.~E.} \bibnamefont{Freedman}},
  \bibinfo{author}{\bibfnamefont{V.~S.} \bibnamefont{Zapf}},
  \bibinfo{author}{\bibfnamefont{N.}~\bibnamefont{Harrison}},
  \bibinfo{author}{\bibfnamefont{D.~G.} \bibnamefont{Nocera}},
  \bibinfo{author}{\bibfnamefont{Y.}~\bibnamefont{Takano}}, \bibnamefont{and}
  \bibinfo{author}{\bibfnamefont{Y.~S.} \bibnamefont{Lee}},
  \bibinfo{note}{arXiv:1402.2693}.

\bibitem[{\citenamefont{Arajs et~al.}(1970)\citenamefont{Arajs, Tehan,
  Anderson, and Stelmach}}]{Febeta}
\bibinfo{author}{\bibfnamefont{S.}~\bibnamefont{Arajs}},
  \bibinfo{author}{\bibfnamefont{B.~L.} \bibnamefont{Tehan}},
  \bibinfo{author}{\bibfnamefont{E.~E.} \bibnamefont{Anderson}},
  \bibnamefont{and} \bibinfo{author}{\bibfnamefont{A.~A.}
  \bibnamefont{Stelmach}}, \bibinfo{journal}{Phys. Lett.}
  \textbf{\bibinfo{volume}{32A}}, \bibinfo{pages}{412} (\bibinfo{year}{1970}).

\bibitem[{Hel()}]{Helton'}
\bibinfo{note}{J. S. Helton, K. Matan, M. P. Shores, E. A. Nytko, B. M.
  Bartlett, Y. Yoshida, Y. Takano, A. Suslov, Y. Qiu, J.-H. Chung, D. G. Nocera
  and Y. S. Lee, Phys. Rev. Lett. \textbf{98}, 107204, (2007)}.

\bibitem[{\citenamefont{Grohol et~al.}(2005)\citenamefont{Grohol, Matan, Cho,
  Lee, Lynn, Nocera, and Lee}}]{Grohol}
\bibinfo{author}{\bibfnamefont{D.}~\bibnamefont{Grohol}},
  \bibinfo{author}{\bibfnamefont{K.}~\bibnamefont{Matan}},
  \bibinfo{author}{\bibfnamefont{J.-H.} \bibnamefont{Cho}},
  \bibinfo{author}{\bibfnamefont{S.-H.} \bibnamefont{Lee}},
  \bibinfo{author}{\bibfnamefont{J.~W.} \bibnamefont{Lynn}},
  \bibinfo{author}{\bibfnamefont{D.~G.} \bibnamefont{Nocera}},
  \bibnamefont{and} \bibinfo{author}{\bibfnamefont{Y.~S.} \bibnamefont{Lee}},
  \bibinfo{journal}{Nat. Mater.} \textbf{\bibinfo{volume}{4}},
  \bibinfo{pages}{323} (\bibinfo{year}{2005}).

\bibitem[{\citenamefont{Matan et~al.}(2010)\citenamefont{Matan, Ono, Fukumoto,
  Sato, Yamaura, Yano, Morita, and Tanaka}}]{MatanVBS}
\bibinfo{author}{\bibfnamefont{K.}~\bibnamefont{Matan}},
  \bibinfo{author}{\bibfnamefont{T.}~\bibnamefont{Ono}},
  \bibinfo{author}{\bibfnamefont{Y.}~\bibnamefont{Fukumoto}},
  \bibinfo{author}{\bibfnamefont{T.~J.} \bibnamefont{Sato}},
  \bibinfo{author}{\bibfnamefont{J.}~\bibnamefont{Yamaura}},
  \bibinfo{author}{\bibfnamefont{M.}~\bibnamefont{Yano}},
  \bibinfo{author}{\bibfnamefont{K.}~\bibnamefont{Morita}}, \bibnamefont{and}
  \bibinfo{author}{\bibfnamefont{H.}~\bibnamefont{Tanaka}},
  \bibinfo{journal}{Nat. Phys.} \textbf{\bibinfo{volume}{6}},
  \bibinfo{pages}{865} (\bibinfo{year}{2010}).

\bibitem[{\citenamefont{Ramirez}(1994)}]{Ramirezf}
\bibinfo{author}{\bibfnamefont{A.~P.} \bibnamefont{Ramirez}},
  \bibinfo{journal}{Annu. Rev. Mater. Sci} \textbf{\bibinfo{volume}{24}},
  \bibinfo{pages}{453} (\bibinfo{year}{1994}).

\bibitem[{\citenamefont{C\"{e}pas et~al.}(2008)\citenamefont{C\"{e}pas, Fong,
  Leung, and Lhuillier}}]{Cepas}
\bibinfo{author}{\bibfnamefont{O.}~\bibnamefont{C\"{e}pas}},
  \bibinfo{author}{\bibfnamefont{C.~M.} \bibnamefont{Fong}},
  \bibinfo{author}{\bibfnamefont{P.~W.} \bibnamefont{Leung}}, \bibnamefont{and}
  \bibinfo{author}{\bibfnamefont{C.}~\bibnamefont{Lhuillier}},
  \bibinfo{journal}{Phys. Rev. B} \textbf{\bibinfo{volume}{78}},
  \bibinfo{pages}{140405(R)} (\bibinfo{year}{2008}).

\bibitem[{\citenamefont{Imai et~al.}(2008)\citenamefont{Imai, Nytko, Bartlett,
  Shores, and Nocera}}]{Imai}
\bibinfo{author}{\bibfnamefont{T.}~\bibnamefont{Imai}},
  \bibinfo{author}{\bibfnamefont{E.~A.} \bibnamefont{Nytko}},
  \bibinfo{author}{\bibfnamefont{B.~M.} \bibnamefont{Bartlett}},
  \bibinfo{author}{\bibfnamefont{M.~P.} \bibnamefont{Shores}},
  \bibnamefont{and} \bibinfo{author}{\bibfnamefont{D.~G.}
  \bibnamefont{Nocera}}, \bibinfo{journal}{Phys. Rev. Lett.}
  \textbf{\bibinfo{volume}{100}}, \bibinfo{pages}{077203}
  (\bibinfo{year}{2008}).

\bibitem[{\citenamefont{Han et~al.}(2012{\natexlab{b}})\citenamefont{Han, Chu,
  and Lee}}]{Han2}
\bibinfo{author}{\bibfnamefont{T.}~\bibnamefont{Han}},
  \bibinfo{author}{\bibfnamefont{S.}~\bibnamefont{Chu}}, \bibnamefont{and}
  \bibinfo{author}{\bibfnamefont{Y.~S.} \bibnamefont{Lee}},
  \bibinfo{journal}{Phys. Rev. Lett.} \textbf{\bibinfo{volume}{108}},
  \bibinfo{pages}{157202} (\bibinfo{year}{2012}{\natexlab{b}}).

\bibitem[{\citenamefont{Mendels and Bert}(2010)}]{MendelsJPSJ}
\bibinfo{author}{\bibfnamefont{P.}~\bibnamefont{Mendels}} \bibnamefont{and}
  \bibinfo{author}{\bibfnamefont{F.}~\bibnamefont{Bert}}, \bibinfo{journal}{J.
  Phys. Soc. Jpn.} \textbf{\bibinfo{volume}{79}}, \bibinfo{pages}{011001}
  (\bibinfo{year}{2010}).

\bibitem[{\citenamefont{Zheng et~al.}(2005)\citenamefont{Zheng, Kawae,
  Kashitani, Li, Tateiwa, Takeda, Yamada, Xu, and Ren}}]{XGZheng}
\bibinfo{author}{\bibfnamefont{X.~G.} \bibnamefont{Zheng}},
  \bibinfo{author}{\bibfnamefont{T.}~\bibnamefont{Kawae}},
  \bibinfo{author}{\bibfnamefont{Y.}~\bibnamefont{Kashitani}},
  \bibinfo{author}{\bibfnamefont{C.~S.} \bibnamefont{Li}},
  \bibinfo{author}{\bibfnamefont{N.}~\bibnamefont{Tateiwa}},
  \bibinfo{author}{\bibfnamefont{K.}~\bibnamefont{Takeda}},
  \bibinfo{author}{\bibfnamefont{H.}~\bibnamefont{Yamada}},
  \bibinfo{author}{\bibfnamefont{C.~N.} \bibnamefont{Xu}}, \bibnamefont{and}
  \bibinfo{author}{\bibfnamefont{Y.}~\bibnamefont{Ren}},
  \bibinfo{journal}{Phys. Rev. B} \textbf{\bibinfo{volume}{71}},
  \bibinfo{pages}{052409} (\bibinfo{year}{2005}).

\bibitem[{\citenamefont{Han et~al.}(2011)\citenamefont{Han, Helton, Chu, Prodi,
  Singh, Mazzoli, M\"{u}ller, Nocera, and Lee}}]{Han}
\bibinfo{author}{\bibfnamefont{T.~H.} \bibnamefont{Han}},
  \bibinfo{author}{\bibfnamefont{J.~S.} \bibnamefont{Helton}},
  \bibinfo{author}{\bibfnamefont{S.}~\bibnamefont{Chu}},
  \bibinfo{author}{\bibfnamefont{A.}~\bibnamefont{Prodi}},
  \bibinfo{author}{\bibfnamefont{D.~K.} \bibnamefont{Singh}},
  \bibinfo{author}{\bibfnamefont{C.}~\bibnamefont{Mazzoli}},
  \bibinfo{author}{\bibfnamefont{P.}~\bibnamefont{M\"{u}ller}},
  \bibinfo{author}{\bibfnamefont{D.~G.} \bibnamefont{Nocera}},
  \bibnamefont{and} \bibinfo{author}{\bibfnamefont{Y.~S.} \bibnamefont{Lee}},
  \bibinfo{journal}{Phys. Rev. B} \textbf{\bibinfo{volume}{83}},
  \bibinfo{pages}{100402R} (\bibinfo{year}{2011}).

\bibitem[{\citenamefont{Mazin et~al.}(2014)\citenamefont{Mazin, Jeschke,
  Lechermann, Lee, Fink, Thomale, and Valenti}}]{Mazin}
\bibinfo{author}{\bibfnamefont{I.~I.} \bibnamefont{Mazin}},
  \bibinfo{author}{\bibfnamefont{H.~O.} \bibnamefont{Jeschke}},
  \bibinfo{author}{\bibfnamefont{F.}~\bibnamefont{Lechermann}},
  \bibinfo{author}{\bibfnamefont{H.}~\bibnamefont{Lee}},
  \bibinfo{author}{\bibfnamefont{M.}~\bibnamefont{Fink}},
  \bibinfo{author}{\bibfnamefont{R.}~\bibnamefont{Thomale}}, \bibnamefont{and}
  \bibinfo{author}{\bibfnamefont{R.}~\bibnamefont{Valenti}},
  \bibinfo{journal}{Nat. Commun.} \textbf{\bibinfo{volume}{5}},
  \bibinfo{pages}{4261} (\bibinfo{year}{2014}).

\bibitem[{\citenamefont{Lee et~al.}(2006)\citenamefont{Lee, Nagaosa, and
  Wen}}]{PALee2006}
\bibinfo{author}{\bibfnamefont{P.~A.} \bibnamefont{Lee}},
  \bibinfo{author}{\bibfnamefont{N.}~\bibnamefont{Nagaosa}}, \bibnamefont{and}
  \bibinfo{author}{\bibfnamefont{X.-G.} \bibnamefont{Wen}},
  \bibinfo{journal}{Rev. Mod. Phys.} \textbf{\bibinfo{volume}{78}},
  \bibinfo{pages}{17} (\bibinfo{year}{2006}).

\end{thebibliography}
\end{document}